\documentstyle[12pt]{article}
\input epsf
\newcommand{\beq}{\begin{eqnarray}}   
\newcommand{\eeq}{\end{eqnarray}}
\newcommand{\ra}{\rightarrow}

\newcommand{\gsim}{\lower.7ex\hbox{$
\;\stackrel{\textstyle>}{\sim}\;$}}
\newcommand{\lsim}{\lower.7ex\hbox{$
\;\stackrel{\textstyle<}{\sim}\;$}}

\begin{document}
\begin{titlepage}
\renewcommand{\thefootnote}{\fnsymbol{footnote}}

\begin{center} \Large
{\bf Theoretical Physics Institute}\\
{\bf University of Minnesota}
\end{center}
\begin{flushright}
TPI-MINN-97/08-T\\
UMN-TH-1540-97\\
ITEP-TH-27/97\\

\end{flushright}
\vspace*{0.7cm}

\begin{center}
{\Large \bf  Domain Walls  in Supersymmetric 
Yang-Mills Theories}
\vspace{0.8cm}

{\Large A. Kovner}

\vspace{0.1cm}

{\it  Theoretical Physics, Oxford Univ. 1 Keble Road, Oxford OX13NP, 
UK$^\dagger$}

 {\it and}

{\it  Theoretical Physics Institute, Univ. of Minnesota,
Minneapolis, MN 55455}

\vspace{0.5cm}

{\Large  M. Shifman} 

\vspace{0.1cm}
{\it  Theoretical Physics Institute, Univ. of Minnesota,
Minneapolis, MN 55455}

\vspace{0.5cm}

{\Large  A. Smilga} 

\vspace{0.1cm}
{\it  Theoretical Physics Institute, Univ. of Minnesota,
Minneapolis, MN 55455}

{\it and}

{\it   Institute of Theoretical and Experimental Physics, Moscow 
117259, Russia$^\dagger$}

\end{center}

\vspace*{.3cm}

\begin{abstract}

We present a detailed analysis of the domain walls in  
supersymmetric gluodynamics and SQCD. We use the (corrected)
Veneziano-Yankielowicz effective  Lagrangians to explicitely obtain 
the 
wall profiles and check  recent   results of Ref. \cite{Dvali1}: (i) the 
BPS-saturated nature of the walls; (ii) the {\em exact} expressions for 
the wall energy density which depend only on global features of 
dynamics
(the existence of a non-trivial  central extension of $N=1$ 
superalgebra
in the theories which admit wall-like solutions).
If supersymmetry is softly broken by the gluino mass, the 
degeneracy of the distinct vacua is gone, and one can consider
the decay rate of the ``false" vacuum into the genuine one.
We do this calculation in the limit of the small gluino mass. 
Finally, we comment on the controversy regarding the existence
of $N$ distinct chirally asymmetric vacua in $SU(N)$ SUSY 
gluodynamics.

\end{abstract}

\vspace{1cm}
\begin{flushleft}

$^\dagger$ Permanent address
\end{flushleft}

\end{titlepage}

\section{Introduction}

Recently it was noted \cite{Dvali1} that some supersymmetric gauge 
theories possess domain walls with rather remarkable properties.
The energy density of these domain walls is exactly calculable, in 
spite of the fact that the theories under consideration are in the 
strong coupling regime. For 
supersymmetric gluodynamics,  the 
 theory of gluons and 
gluinos with no matter, 
the calculation of the energy density was carried out in Ref. 
\cite{Dvali1}, in an indirect way.  The key ingredient is the
central extension of the $N=1$ superalgebra,
\beq
\{ Q^\dagger_{\dot\alpha}Q^\dagger_{\dot\beta}\}
=
\frac{N}{4\pi^2}\left(\vec\sigma\right)_{\dot\alpha\dot\beta}\int
 \, d^3 x \, \vec\nabla\left( \mbox{Tr}\, \lambda^2
\right) \, ,
\label{cext}
\eeq
where $Q^\dagger_{\dot\alpha}$ is the supercharge,
$\lambda$ is the gluino field, and  
$\left(\vec\sigma\right)_{\dot\alpha\dot\beta}
=\{ \sigma^3, -i , -\sigma^1\}_{\dot\alpha\dot\beta}
$ is a set of matrices converting the vectorial index
of the representation $(1,0)$ of the Lorentz group in the spinorial
indices 
\footnote{Equations (9) and (11)  
in the original version   of Ref. \cite{Dvali1} contained 
a misprint, the factor $N/4\pi^2$ was omitted. Further comments 
regarding Eq. (\ref{cext}) are presented in Sect. 4.}.  The commutator 
(\ref{cext}) is given for $SU(N)$ gauge group;
the parameter $N$  reflects this choice of the group. 
The integral over the full derivative on the right-hand side
is zero for all localized field configurations; it does not vanish, 
however, for the domain walls. Equation (\ref{cext}) implies that the 
energy density of the domain wall is
\beq
\varepsilon = \frac{N}{8\pi^2}
\left|\langle \mbox{Tr}\, \lambda^2\rangle_\infty
-\langle \mbox{Tr}\, \lambda^2\rangle_{-\infty}
\right|\, ,
\label{vacen}
\eeq
where the subscript $\pm\infty$ marks the values of the gluino 
condensate at spatial infinities (say, at $z\ra\pm\infty$ assuming 
that the domain wall lies in the $xy$ plane). The existence of the 
exact relation (\ref{vacen}) is a consequence of the fact that
the domain wall in the case at hand is a BPS-saturated configuration
preseving $1/2$ of the original supersymmetry.

In this paper we will explore in more detail the issue of the domain 
walls both in supersymmetric gluodynamics (Sect. 2) and in 
supersymmetric 
extension of QCD (SQCD, supersymmetric Yang-Mills theory with 
matter), see Sect. 3, 4 and 5.  
We will consider the profiles of the domain wall solutions, 
and calculate the energy density directly, by analyzing these profiles. 
The expressions obtained in this way will be confronted with the 
general results of Ref. \cite{Dvali1}.  Another issue of interest,
to be discussed below, is the dependence of the central charge
on the mass parameter of the matter field $m_0$. The central charge 
is a chiral quantity; therefore, the dependence on $m_0$ should
be holomorphic, as in Ref. \cite{Shif1}. The holomorphy implies, that
as far as the energy density of a BPS--saturated domain wall
 is concerned, the transition from
the weak coupling Higgs regime to the strong coupling 
supersymmetric gluodynamics is smooth.

If supersymmetry is explicitly (softly) broken, say, by the
gluino mass term, the vacuum degeneracy is lifted -- we find 
ourselves in a classical situation with false vacuum.
Now, instead of the domain wall, one can study the decay rate of the
false vacuum. If supersymmetry breaking is small, so that
it is legitimate to  work in the leading order in this parameter, 
 one can obtain an explicit expression for the decay rate of the
false vacuum. This problem is discussed in Sect. 6. 

Section 7 is devoted to
an issue which, although related to the domain wall solutions, 
can be formulated in wider terms. Questioned is the very existence of 
$N$ distinct chirally asymmetric vacua in the $SU(N)$
supersymmetric gluodynamics. The chiral $Z_{2N}$ symmetry is a 
remnant of the anomalous $R_0$ symmetry of the model.
The presence of this symmetry is due to quantization of the 
topological charge. If the topological charge is quantized in a 
non-standard way (i.e. fractional topological charges are allowed
\footnote{Another way to make the same statement is to say
that non-contractable cycles in the space of the gauge fields are 
shorter than it is usually believed.}) then the global structure of the
theory changes. In particular, only one chirally asymmetric vacuum 
survives, not $N$. The extra states disappear as a result of a new 
superselection rule. This is clearly seen within the effective
Lagrangian approach. The issue is being debated in the literature.
We discuss the impact of new arguments associated with 
supersymmetry
and smooth (and calculable) behavior of the wall energy density.

\section{Supersymmetric Gluodynamics}

To begin with, we will concentrate   on supersymmetric 
generalization
of pure gluodynamics -- i.e. the theory of gluons and gluinos. The 
Lagrangian of the model at the fundamental level has the form 
\cite{FZ}
\beq
{\cal L} =  -\frac{1}{4g_0^2} 
G_{\mu\nu}^aG_{\mu\nu}^a + {\vartheta\over 
32\pi^2}G_{\mu\nu}^a\tilde G_{\mu\nu}^a
+\frac{1}{g_0^2}\left[
i\lambda^{a \alpha} 
D_{\alpha\dot\beta}\bar\lambda^{a\dot\beta}
\right] \, 
,
\label{SUSYML}
\eeq
where the spinorial notation is used. In the superfield language
the Lagrangian can be written as
\beq
{\cal L} =  \frac{1}{4g^2} \mbox{Tr} \int d^2\theta W^2 \ + {\rm 
H.c.}\, ,
\label{SFYML}
\eeq
where
$$
\frac{1}{g^2} = \frac{1}{g_0^2} -\frac{i\vartheta}{8\pi^2}\, .
$$
In what follows the vacuum angle $\vartheta$ will play no role and 
can be set 
equal to zero.  Our conventions regarding the superfield formalism 
are summarized e.g. in the recent review
\cite{Shif2}. We will limit ourself to the $SU(N)$ gauge group
(the generators of the group $T^a$ are in the fundamental 
representation, so that $\mbox{Tr}(T^aT^b) = (1/2)\delta^{ab}$).

$SU(N)$ supersymmetric gluodynamics 
has a discrete symmetry,  $Z_{2N}$,   a (non-anomalous) 
remnant of the 
anomalous axial 
symmetry generated by the phase rotations of the gluino field. The 
formation of the 
gluino condensate $\langle\lambda \lambda\rangle$ (it follows from
certain supersymmetric Ward identities combined with explicit 
instanton
calculations \cite{lamlam}) breaks this
 discrete chiral 
symmetry down to $Z_2$.
Therefore, there exists a set of distinct vacua labelled by the
value of the gluino condensate. The field configurations interpolating 
between different values of $\langle\lambda \lambda\rangle$ 
at spatial infinities are topologically stable domain walls.

A formal description of these domain walls can be given
in the framework of the effective Lagrangian approach.
We will exploit  the so-called Veneziano-Yankielowicz (VY) effective 
Lagrangians \cite{VY,TVY}.  The original VY Lagrangian
does not possess \cite{Vene} the discrete $Z_{N}$ invariance of
supersymmetric gluodynamics (\ref{SUSYML}). Recently it 
was 
shown  that the VY expression is  
 incomplete;  it was  amended to become 
compatible with all symmetries of supersymmetric gluodynamics
in Ref. \cite{Kovn1}.
The corrected expression exhibits $N$ minima of the scalar 
potential corresponding to $Z_{2N}\ra Z_2$ breaking,
{\it plus}  an additional  minimum at the origin
where the gluino condensate vanishes. 

For simplicity from now on, if not stated to the contrary, we
will consider the case of $SU(2)$, although this restriction is not
of any conceptual importance and can be easily lifted.

The Lagrangian realizing the anomalous Ward
identities is  constructed in terms of 
the chiral superfield 
\beq
S = \frac{3}{32\pi^2}\,\mbox{Tr}\,W^2\, ,
\label{superfs}
\eeq
 namely
\beq
{\cal L} = \frac{1}{4}\int d^4\theta C
\left( \bar S S \right)^{1/3} + 
\frac{1}{3} \int d^2\theta S\left( \ln \frac{S^2}{\sigma^2} +2\pi i 
n\right) + 
\frac{1}{3} \int d^2\bar\theta \bar S\left(  \ln 
\frac{\bar S^2}{\bar\sigma^2} - 2\pi i n\right) 
 ,
\label{VYL}
\eeq
where 
$\sigma$ is a numerical parameter,
$$
\sigma = {\rm e}\Lambda^3 {\rm e}^{i\vartheta /2}\, ,
$$
$\Lambda$ is the  scale parameter, a positive number of dimension 
of mass which we will set equal to unity in the following.

A new element in the Lagrangian (\ref{VYL}) is an integer-valued  
Lagrange multiplier  $n$. In calculating the partition function and all 
correlation functions the sum over $n$ is implied. 
The variable $n$ takes only integer values and is not a local
field. It does not depend on the space-time coordinates and, 
therefore,
integration over it 
imposes  a global constraint on the 
topological 
charge. It is easy to see that (after the Euclidean rotation) the 
constraint  takes the form
\beq
\label{topch}
\nu \ =\ \frac{1}{32\pi^2}\int d^4x G_{\mu\nu}^a\tilde G_{\mu\nu}^a 
= Z\, .
 \eeq

While the $F$ term in Eq. (\ref{VYL}) is unambiguously fixed, the 
$D$ term
is not specified by the anomalous Ward identities.
First, the effective Lagrangian at hand is not Wilsonean;
therefore,  there are no reasons to discard terms with higher 
derivatives, generally speaking.  A possible example of this
type is
\beq
\left. \frac{(\partial_\mu \bar S^{1/3}\partial_\mu S^{1/3})}{(\bar 
SS)^{1/3}}\right|_D \, .
\eeq
Even leaving aside higher derivatives,  one is free to choose any
value of the numerical constant $C$ in Eq. (\ref{VYL}).  As we will 
see shortly, these ambiguities do have an impact on the profile of the 
wall. The surface energy density, however, remains intact, in full 
accord with
the general arguments of Ref. \cite{Dvali1}. For the time being we 
will put $C=1$. 

The extra term 
added to the Veneziano - Yankielowicz
Lagrangian is clearly supersymmetric  and is also 
invariant
under all global symmetries of the original theory. The 
single-valuedness of the scalar potential and the $Z_2$ 
invariance
are restored \footnote{The 
explicit invariance here 
is $Z_2$ rather than the complete $Z_{4}$ of the original
SUSY gluodynamics, since we have chosen to write our effective 
Lagrangian for 
the superfield which is invariant under $\lambda\rightarrow 
-\lambda$.}.
The chiral phase rotation by the angle $ \pi k$ with 
integer
$k$ just leads to the shift of $n$ by $k$ units. Since $n$ is summed 
over in the functional integral,
the resulting Lagrangian for $S$ is indeed $Z_2$ 
invariant.
  
The constraint on $[S-\bar S]_F$ following from the Lagrangian 
(\ref{VYL}) results in a peculiar form of the 
scalar potential. 
The expression for the scalar potential
is given in Eq. (13) of Ref. \cite{Kovn1}.
It can be considerably simplified in the infinite volume limit.
Eliminating, as usual the $F$ component of $S$ with the help of 
classical 
equations of motion at fixed $n$, the effective potential can be 
written as
\beq
U(\phi)\ =\ -V^{-1}\ln\left[
\sum_n\exp\{-16 V (\phi^*\phi)^{2/3}[\ln^2|\phi|+(\alpha +\pi 
n)^2]\}\right]\, .
\label{scalpot}
\eeq
Here $V$ is the total space-time volume of the system, $\phi$
is the lowest component of the superfield $S$, and 
$\alpha={\rm arg}
 (\phi)$.
In the limit $V\rightarrow\infty$ only one term in the sum over $n$ 
contributes
for every value of $\alpha$. Thus, for $-\pi/2<\alpha<\pi/2$ the only 
contribution comes from $n=0$, while for $\pi/2<\alpha<3\pi/2$ 
from $n=-1$.
Therefore in the right half plane
\beq
U(\phi ) \equiv U_0 (\phi )= 16(\phi^*\phi)^{2/3}
\ln \phi \ln \phi^*
\,\,\mbox{at arg}\phi\in (-
\frac{\pi}{2}, \frac{\pi}{2})\, ,
\label{Uphi}
\eeq
while in the left half-plane, when arg$\phi \in (\pi /2, 3\pi /2 )$ 
\beq
U(\phi ) = U_0 (\phi e^{-i\pi})\, .
\label{Umphi}
\eeq
In other words, the complex $\phi$ plane is divided into two 
sectors. The scalar potential in the second sector is just that in the 
first sector rotated by $-\pi$. The scalar potential itself is continuous,
but its first derivative in the angular direction experiences a jump
at arg$\phi = \pm \pi/2$. The scalar potential is ``glued" out of two 
pieces
\footnote{Similar ``glued'' potentials appear due to the quantization
of topological charge in the Schwinger model \cite{QCD2}.}.
The $Z_2$ symmetry is explicit in this expression.
It is quite obvious that the problem at hand has three 
supersymmetric minima -- two at $\phi = \pm 1$, corresponding 
to a non-vanishing value of the gluino condensate (spontaneously 
broken discrete chiral symmetry), and a minimum at $\phi = 0$
(unbroken chiral symmetry)\footnote{The existence of this 
{\it additional} vacuum state with vanishing gluino
condensate, which does not follow from any symmetry 
considerations
is probably the most surprising and nontrivial finding of Ref. 
\cite{Kovn1}
and is also a very important element of the present study of domain 
walls.
We want to draw an analogy here with the situation
in two--dimensional QCD with adjoint fermion studied in Ref. 
\cite{QCD2}.
There too vacuum structure is nontrivial and ``domain walls'' 
interpolating
between the vacua with different values of the fermion condensate 
exist.
For $SU(N)$ gauge groups with even $N \geq 4$, the existence of 
different
vacua did not follow from symmetry considerations. 
In Ref. \cite{QCD2}, this
was formulated as a paradox. It seems now that it is not a logical 
paradox,
but rather a surprising phenomenon which takes place in some 
sophisticated
enough two--dimensional and four--dimensional gauge theories.}.

\section{Domain Walls in Supersymmetric Gluodynamics}

 We will now consider the 
domain wall interpolating between 
$$
\phi = 0 \ \ \ {\rm and} \ \ \ \phi = 1
$$
 at spatial infinities. 

 For definiteness, the domain wall
is assumed to be in the $xy$ plane. The profile to be 
determined is a function of one coordinate, $z$.
The corresponding domain wall is purely real (Im$\phi$
will vanish on the wall profile) and lies completely inside the first 
sector.  In other words, we will 
only need the potential on the real positive semiaxis of  $\phi$.

It is convenient to perform a rescaling 
$$
S = \Phi^3
$$
which casts the kinetic term of the scalar field in the canonic form.
Correspondinly,
$$
\phi=\varphi^3\, ,
$$
where $\varphi$ is the lowest component of the superfield $\Phi$.

The scalar potential in the right half plane of $\phi$ then is 
\beq
U(\varphi ) = \left| \frac{\partial{\cal W} 
(\varphi )}{\partial\varphi}\right|^2
\eeq
where
\beq
{\cal W} (\varphi ) = \frac{2}{3} 
\varphi^3\ln\frac{\varphi^6}{\mbox{e}^2} \, .
\label{supot}
\eeq

The domain wall is the planar static field configuration
$\varphi (z)$ satisfying the boundary conditions $\varphi(-\infty) = 
0,
\varphi(\infty) = 1$ and minimizing the energy functional. Hence,
$\varphi (z)$ satisfies the classical equations of motion with the 
inversed
sign of the  potential:

\beq
\partial^2_z \varphi = \frac{\partial^2\bar{\cal 
W}}{\partial\bar\varphi^2}
\frac{\partial{\cal W}}{\partial\varphi}\, .
\label{geneq}
\eeq
As we will now show the domain wall we deal with is a 
BPS-saturated state \cite{Dvali1}.
This means that a linear combination of supercharges, acting on the 
wall, annihilates it. One half of supersymmetry is preserved. 
The general formula for the supersymmetry transformation is
\beq
\delta\psi^\alpha =\sqrt{2} i (\partial^{\dot\beta\alpha}\varphi 
)\bar\epsilon_{\dot\beta} + \sqrt{2}\epsilon^\alpha F\, , \,\,\,
\delta\bar\psi^{\dot\alpha} =- \sqrt{2} i 
(\partial^{\dot\alpha\alpha}\bar\varphi )\epsilon_{\alpha} + 
\sqrt{2}\bar\epsilon^{\dot\alpha}\bar{F}\, .
\label{supertra}
\eeq
If the parameter of the transformation $\epsilon^\alpha$ is chosen in 
such a way that
$$
\sigma_1 \epsilon^{*} = \pm i\epsilon\, ,
$$
then $\delta\psi =0$ provided that 
\beq
\partial_z \varphi = \pm F\, ,
\label{BPSeq}
\eeq
where $F = -\partial\bar{\cal W}/\partial\bar\varphi$. 

Equation (\ref{BPSeq}) is the first-order differential equation,
a ``square root" of the general equation (\ref{geneq}). The solution 
with the boundary conditions we are interested in
corresponds to the plus sign in Eq. (\ref{BPSeq})
\footnote{The correponding supertransformation parameter
$\epsilon^\alpha$ is 
$\epsilon^{1,2 }= \{1,-i\}$.}; under our choice of parameters it is 
obviously purely 
real. Therefore, $\bar{\cal W}={\cal W}$.
Even without knowing the explicit form of the solution, the energy 
density
of the wall can be readily calculated. Indeed,
\beq
\varepsilon = \int_{-\infty}^{\infty}dz\, 
\left[ \partial_z\bar\varphi \partial_z \varphi 
+\frac{\partial{\cal W}}{\partial \varphi}\frac{\partial\bar{\cal 
W}}{\partial\bar\varphi}
\right] =
2\int_{-\infty}^{\infty}dz\, \left(\partial_z\bar\varphi  \right)\left( 
-\frac{\partial\bar{\cal W}}{\partial\bar\varphi}
\right)\, ,
\label{exeqe}
\eeq
where Eq. (\ref{BPSeq}) is used. The right-hand side evidently 
reduces to
\beq
\varepsilon = 2\left(\bar{\cal W}_{-\infty} -  \bar{\cal W}_{\infty} 
\right) = \frac{8}{3} \, ,
\label{wed}
\eeq
where the values of the superpotential (\ref{supot}) at
$z =\pm\infty$ (i.e. $\varphi = 0$ and $1$) are substituted.
This result is in full agreement with Eq. (\ref{vacen}) \footnote{We 
remind that $\Lambda$ is set equal to unity, so that $3(32\pi^2)^{-
1}$Tr$\lambda^2=1$ in the vacuum with the broken chiral
invariance.}.

Let us discuss now the wall profile.  Combining Eqs. (\ref{supot}) and 
(\ref{BPSeq}) we arrive at the following relation
\beq
\int^2_{1/\varphi} \frac{dx}{\ln x} = 12 (z -z_0)\, ,
\label{profeq}
\eeq
where $z_0$ is the wall center. The left-hand side is expressible in 
terms of the integral logarithm. This does not help us find
the explicit form $\varphi (z)$ and we therefore will not use this 
expression
in the following.
The asymptotic behavior of the solution is however
transparent. At large positive $z$,
$\varphi (z)$ approaches unity exponentially,
$$
\varphi (z)\ra 1 - \mbox{Const}\, e^{-12(z-z_0)}\, .
$$
At large negative $z$, $\varphi (z)$ approaches zero as
$$
\varphi  \sim \frac{1}{12} \frac{1}{z_0 - z}
$$
modulo logarithms. This type of behavior was anticipated.
Indeed, for positive $z$ we are in the phase with the spontaneous
breaking of the chiral symmetry and a mass gap.
Hence, the approach is exponential. On the other hand,
the phase at $z\ra -\infty$ has no mass gap
\cite{Kovn1}, and the asymptotics is power-like. 

Figure 1 presents the profile of the domain wall for the
field $\varphi (z)$.
\begin{figure}
\epsfbox{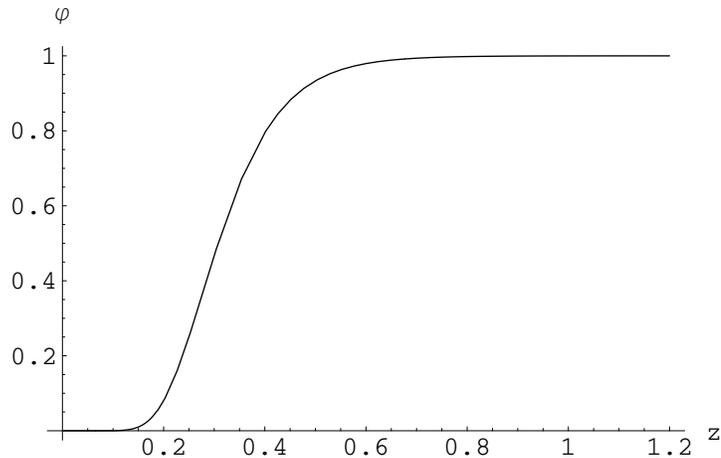}
\caption{ The domain wall profile in SUSY gluodynamics.}
\label{f1}
\end{figure}

Note that the
wall energy density calculated above is insensitive to the 
particular choice of the kinetic term. Eq. (\ref{exeqe})
illustrates this insensitivity. Let us return to Eq. (\ref{VYL})
and restore the constant $C$ in the kinetic term, $C\neq 1$,
as an example of possible ambiguity. It is quite clear that
the constant $C$ is immediately absorbed in  a rescaling
of $z$, and the final expression   (\ref{wed}) remains intact. 

\section{SQCD with One Flavor}

What happens with the domain walls when the matter fields are 
added?
In this section we will consider $SU(2)$ theory with one flavor
(two subflavors). The matter fields belong to the fundamental
representation of $SU(2)$. This model is described in great detail in 
the review paper \cite{VZS}, so we omit all explanations.
At the fundamental level the matter sector has the form
\beq
{\cal L}_M = \frac{1}{4}\int d^2\theta d^2\bar\theta^2
\bar Q^f  e^V Q^f +\left( \frac{m_0}{4} \int d^2\theta 
Q^{\alpha f}Q_{\alpha f} +\mbox{H.c}\right)\, ,
\label{matter}
\eeq
where $\alpha$ is the color and $f$ the subflavor index, $\alpha , f = 
1,2$; $Q$ is the quark superfield. Furthermore, $m_0$ is the matter 
mass term. The subscript
0 indicates that it is the bare mass  that enters the Lagrangian;
this parameter is complex. Certain quantities depend on $m_0$
in a holomorphic way, which will allow us to exactly trace
the evolution of the wall parameters under variations of  $m_0$ 
\cite{Shif1,Shif3}. 

By examining the symmetries of the model one  derives the 
corresponding VY effective Lagrangian \cite{TVY}.  The $R_0$ 
current, the superpartner of the energy-momentum tensor and 
supercurrent \cite{FZ2}, generates
\beq
\lambda_\alpha \ra e^{i\beta}\lambda_\alpha\, , \,\,\, 
\psi_\alpha^f \ra e^{-(i/3)\beta} \psi_\alpha^f \, , \,\,\, 
\phi_\alpha^f \ra e^{(2i/3)\beta} \phi_\alpha^f \, ,
\label{rnot}
\eeq
where $\psi$ and $\phi$ are the quark and squark fields, 
respectively. The $R_0$ 
current is anomalous at the quantum level. The anomaly free $R$
current is a linear combination of the $R_0$  current and the Konishi
current \cite{Koni1}. At the one-loop level it corresponds to 
\cite{ADS}
\footnote{In higher orders the form of the conserved $R$ current 
becomes more complicated \cite{KSV}, but this is unimportant for 
our purposes.}
\beq
\lambda_\alpha \ra e^{i\beta}\lambda_\alpha\, , \,\,\, 
\psi_\alpha^f \ra e^{-2i\beta} \psi_\alpha^f \, , \,\,\, 
\phi_\alpha^f \ra e^{-i\beta} \phi_\alpha^f \, .
\label{rcur}
\eeq
The $R$ current is conserved in the massless limit $m_0=0$.

The VY Lagrangian is constructed in terms of two chiral superfields, 
$S$ 
(see
Eq. (\ref{superfs})) and $M$,
\beq
M = Q^{\alpha f}Q_{\alpha f}\, .
\label{superfm}
\eeq
Omitting irrelevant constants one can write
$$
{\cal L} = \frac{1}{4}\int d^4\theta 
\left( \bar S S \right)^{1/3} + \frac{1}{4}\int d^4\theta 
\left( \bar M M \right)^{1/2} +
$$
\beq\left[
\frac{1}{3} \int d^2\theta S\left( \ln \frac{SM}{e\tilde\Lambda^5} 
+2\pi i 
n\right) 
+\frac{m}{4}\int d^2\theta M +\mbox{H.c.}
\right]\,  .
\label{VYM}
\eeq
Following Ref. \cite{Kovn1}, we introduced the Lagrange multiplier
(integer-valued constant ``field" $n$)
in the original expression which can be borrowed e.g. from
\cite{Amati}, to ensure proper quantization of the topological charge.
As before, summation over $n$ is implied. {\em A priori} the mass 
parameter (the coefficient in front of $M_F$)
is proportional to the quark mass $m_0$. The coefficient of 
proportionality ($1$) was  established from the Konishi anomaly 
\cite{Koni1}, see below. Furthermore, $\tilde\Lambda$
is the scale parameter of SQCD with one flavor. Its relation to 
$\Lambda$, the scale parameter of supersymmetric gluodynamics 
will be established later. 
Again, as in supersymmetric gluodynamics, the $D$ terms in the 
Lagrangian  (\ref{VYM}) are not determined completely by the 
symmetries
of the theory. We have chosen the simplest $D$ term which is as 
good
as any other one for the purpose of illustrating our point.

In the massless limit, $m_0\ra 0$, the Lagrangian (\ref{VYM}) is 
obviously invariant
under the $R$ transformation, $\theta\ra e^{i\beta}\theta$,
$S\ra e^{2i\beta}S$ and $M\ra e^{-2i\beta}M$. The variation of the 
 Lagrangian (\ref{VYM}) under the anomalous $R_0$ transformation
is 
$$
\frac{20}{9} i\beta (S-\bar S)_F\, ,
$$
while that of (\ref{VYL}) is
$$
\frac{8}{3} i\beta (S-\bar S)_F\, ,
$$
in full accord with the fact that the first coefficients of the $\beta$
functions are $5$ and $6$ in $SU(2)$
 SQCD with one flavor and supersymmetric gluodynamics, 
respectively.  If $m_0\neq 0$ the only chiral invariance left in the 
Lagrangian (\ref{VYL}) is the discrete (anomaly free) $Z_4$ chiral 
transformation
$$
\lambda_\alpha \ra e^{\pi i k/2}\lambda_\alpha\, , \,\,\, 
\psi_\alpha^f \ra  \psi_\alpha^f \, , \,\,\, 
\phi_\alpha^f \ra e^{\pi i k/2} \phi_\alpha^f \, ,
$$
where $k$ is an integer. 
In the VY Lagrangian (\ref{VYM}), amended in accordance with Ref.
\cite{Kovn1},  the discrete chiral symmetry  is realized as $Z_2$,
$$
\theta\ra e^{\pi i k/2} \theta\, , \,\,\, 
S\ra e^{\pi i k} S \, , \,\,\, M\ra e^{\pi i k} M\, .
$$
The scalar potential, being properly calculated from Eq. (\ref{VYM}),  
is composed of  two sectors, much in the same way 
as in supersymmetric gluodynamics, Sect. 2. 

The fundamental theory described by the Lagrangian
(\ref{SUSYML}),  (\ref{matter}) is in the unified Higgs/confinement 
phase, since the Higgs field is in the fundamental representation.
One can distinguish the strong coupling regime versus the weak 
coupling regime. If the expectation value of the field $M$ is small
(this happens either for large values of $m_0$ or in
the additional vacuum found in Ref. \cite{Kovn1}) we are in the 
strong coupling regime, where the VY effective Lagrangian is 
expected
to properly capture the qualitative picture of the emerging dynamics.
If the expectation value of the field $M$ is large, we are in the weak 
coupling regime \cite{ADS}. Here the VY Lagrangian provides a good 
description
of the moduli fields $M$, but
does not properly describe the dynamics 
of the massive $W$ bosons characteristic of the weak coupling 
regime. Yet, it reproduces correctly the vacuum structure in both 
cases. Correspondingly, the profile of the 
domain wall
following from Eq. (\ref{VYM}) will be qualitatively realistic. 
In this section we will consider the wall solution interpolating
between $\phi=1$ and $\phi=0$.
The solution that interpolates between two chirally noninvariant
vacua at small $m_0$ will be treated separately in Sect. 5. As in
supersymmetric gluodynamics, the energy density of the walls will 
be exact.

Again, we will be interested here in a solution interpolating between
the vacuum at the origin and that at a finite values of $S,M$.
Therefore, it is sufficient to consider only one sector
of the theory. 
The corresponding vacuum structure is obtained by minimizing the 
superpotential
\beq
{\cal W} = \frac{2}{3} S\ln \frac{SM}{{\rm e}\tilde\Lambda^5} + 
\frac{m}{2} M \,  .
\label{sptm}
\eeq

 From $\partial {\cal W}/\partial M = 0$ we conclude that
in the standard vacuum
\beq
\langle S \rangle = -\frac{3}{4} m\langle   M \rangle \, .
\label{comko}
\eeq
Given our definitions of $S$ and $M$, this is nothing but 
a consequence of the Konishi relation \cite{Koni1},
\beq
\bar D^2 \bar Q^{\alpha f} e^V Q^{\alpha f} = 4
Q^f\frac{\partial{\cal W}}{\partial Q^f} + \frac{1}{2\pi^2}
\mbox{Tr} W^2\ =
4m_0  Q^{\alpha f}  Q_{\alpha f} +\frac{1}{2\pi^2}
\mbox{Tr} W^2\, .
\label{konishi}
\eeq
This explains our choice of the mass parameter in Eq. (\ref{VYM}).

The condition $\partial {\cal W}/\partial S = 0$ implies that
 the vacuum states are at 
$\langle S \rangle \langle   M \rangle  =1$ (in the units
of $\tilde\Lambda$ which will be used hereafter, if not stated to the 
contrary).  It is seen that it is convenient to assume the mass 
parameter $m$ to be  real  and negative. Then the vacuum value of 
$M$ will be real. The phase of the parameter $m$
can be adjusted arbitrarily by an appropriate rotations of the fields
$S$ and $M$.  From now on we will assume that $m$ is  a real 
negative number; for convenience we will introduce 
$$
\tilde m = - m\, .
$$
Then $\tilde m$ is real and positive. 

It is convenient to pass
to the superfields with the canonic kinetic terms,
\beq
S \ra \Phi^3\, , \,\,\, M\ra  X^2\, .
\label{cnkt}
\eeq
The lowest component of $\Phi$ is denoted by $\varphi$, as in Sect. 
2; the lowest component of $X$ is $\chi$.
The superpotential (\ref{sptm}) generates the scalar potential
 \beq
U(\varphi, \chi) \ =\ \left|\frac{\partial W}{\partial \varphi}\right|^2 
+
 \left|\frac{\partial W}{\partial \chi}\right|^2 \ =\ 
4\left| \varphi^2 \ln(\varphi^3 \chi^2) \right|^2 +
\left|\chi\left(\tilde{m} - \frac{4\varphi^3}{3\chi^2} \right)\right|^2
\label{potfihi}
  \eeq
It is zero at the origin and at the standard minima (\ref{comko}).
 
 Let us analyze the profile of the domain wall connecting the 
minimum at
the origin and one of the standard minima (with the positive sign of
$\langle M\rangle = \langle \chi^2\rangle $). It
is again a  BPS-saturated state, preserving two out of four 
supercharges,
see Eq. (\ref{supertra}). The corresponding equations are
\beq
\partial_z \varphi = \pm F_\Phi\, ,\,\,\,  \partial_z \chi
= \pm F_X\,
,
\label{eqdw}
\eeq
where
$$
\bar F_\Phi = - \frac{\partial{\cal W} (\varphi , \chi 
)}{\partial\varphi }\, , \,\,\, 
\bar F_X = - \frac{\partial{\cal W} (\varphi , \chi )}{\partial\chi }
\, .
$$
The following boundary conditions are imposed:
  \beq
\label{bcdw}
\varphi =\chi = 0\,\,\,
\mbox{at}\,\,\,  z\ra-\infty ;\,\,\, \varphi \ra \varphi_*\, ,\,\, 
\chi\ra\chi_*\,\,\,
\mbox{at}\,\,\,  z\ra\infty\, ,
  \eeq
where (the real positive) parameters $\varphi_* ,\chi_*$
are defined through the expressions
$$
\chi_*^2 =\frac{2}{\sqrt{3}}\frac{1}{\sqrt{\tilde{m}}}\, ,\,\,
\varphi^3_* =\frac{\sqrt{3}}{2}\sqrt{\tilde{m}}\, ,
$$
and the scale parameter $\tilde\Lambda$ is put to unity, 
temporarily. These boundary conditions correspond to the plus sign 
in
Eq. (\ref{eqdw}). 
After the rescaling $\chi = \tilde{m}^{-1/4} \tilde \chi,\,\,\, 
 \varphi = \tilde{m}^{1/6} \tilde \varphi$, Eqs. (\ref{eqdw})
take the form
  \beq
 \label{eqdwsc}
\partial_z \tilde \varphi \ =\ -2\tilde{m}^{1/6} \tilde \varphi^2 
\ln(\tilde \varphi^3 \tilde \chi^2)\, , 
\nonumber \\
\partial_z \tilde \chi \ =\  \tilde{m}\tilde\chi\left(1 - 
\frac{4\tilde\varphi^3}
{3\tilde\chi^2}\right)\, .
 \eeq
These equations can be solved analytically in two interesting cases,
 $\tilde{m} \to \infty$ and $\tilde{m} \to 0$.

When the mass is large, the behavior of the solution is qualitatively 
evident.
In this case, we can integrate out the heavy matter fields after which 
the 
dynamics of the light (gauge) sector is the same as in 
supersymmetric 
gluodynamics. Indeed, 
 at $z =\infty$ the curve starts at $\varphi_*,
\chi_*$, then it follows the trajectory on which $\varphi^3$
is (almost) equal to $3\tilde{m}\chi^2/4$, and the solution for 
$\varphi$
is (almost) the same as in Sect. 2 -- exponentially close to 
$\varphi_*$, then quickly $\varphi$ and $\chi$ approach zero
as $(z_0 - z)^{-1}$ modulo logarithms. The condition $\varphi^3 = 
3\tilde{m}\chi^2/4$
nullifies the (large) second term in the scalar potential
(\ref{potfihi}) and exactly corresponds to freezing the heavy 
degree of
freedom in the spirit of the Born-Oppenheimer approach.

The situation is somewhat nontrivial in the opposite limit when the
matter fields are very light. A standard procedure would be to freeze 
the gauge degrees of freedom which amounts to imposing the 
condition
 $\varphi^3 \chi^2 = 1$ (so that the ``large'' first term in the scalar 
potential
would disappear). After this, the effective potential for the light 
matter fields
is obtained \cite{ADS} which gives a valid description 
of the ``Higgs'' phase,
$|\chi| \gg 1$. We will analyze in more detail the physics of the 
Higgs phase
in the subsequent section and will show, in particular, that BPS--
saturated
domain walls which interpolate between two Higgs vacua 
(\ref{comko})
with different phases of $\chi$ do 
exist. However, the domain walls interpolating
between one of 
the standard vacua and the new vacuum at the origin disappear in 
this
approach.

We want to emphasize that Eqs. (\ref{eqdw}),  (\ref{eqdwsc}) 
with the
boundary
conditions (\ref{bcdw}) {\it do} have a solution at any value of 
$\tilde{m}$. 
In the
limit $\tilde{m} \to 0$, the solution acquires, however, a peculiar 
singular 
form
being composed of two pieces. The first piece corresponds to moving 
from the
minimum $\varphi = \chi = 0$ at $z = -\infty$ to the point $\varphi 
= 0,
\chi = \chi_*$ at $z = z_0$ according to the law  
 $$ 
 \chi(z) \ =\ \chi_* e^{\tilde{m}(z-z_0)} \, .
 $$
Then the trajectory abruptly turns: $\chi(z)$ does not change 
anymore, but
 $\varphi(z)$ starts  rising from $\varphi(z_0) = 0$ to 
$\varphi(\infty) =
\varphi_*$ as is dictated by the first equation in (\ref{eqdwsc}) with 
frozen
$\chi(z) = \chi_*$ (it has the same functional form as the wall 
equation in
supersymmetric gluodynamics). 
This second half of the wall is much broader 
than 
the first
one. Its width is of order $\tilde{m}^{-1/6}$ compared to the width 
$\sim 
1/{\tilde{m}}$ of
the section of the wall with negative $z-z_0$.

It is clear why this solution is lost in the Born-Oppenheimer 
analysis. The
scalar potential (\ref{potfihi}) involves a rather high barrier between 
the
 Higgs
vacua and the origin.  The new wall corresponds to
climbing this ``mountain ridge''. 
In the vicinity of the ridge the relation $\varphi^3\chi^2=1$
is not valid and the naive Born-Oppenheimer analysis breaks down.

It is not surprising as such
that a solution going
over this ridge still exists. What is rather remarkable and very 
specific for
supersymmetric case is that the surface energy density of this 
non-standard
BPS--saturated wall is not high.
 It is determined by the generalization of Eq. 
(\ref{exeqe}),
\beq
\varepsilon = \int_{-\infty}^{\infty}dz\, 
\left[ \partial_z\bar\varphi \partial_z \varphi +
\partial_z\bar\chi \partial_z \chi 
+\frac{\partial{\cal W}}{\partial \varphi}\frac{\partial\bar{\cal 
W}}{\partial\bar\varphi} 
+\frac{\partial{\cal W}}{\partial \chi}\frac{\partial\bar{\cal 
W}}{\partial\bar\chi}
\right] =
2\left(\bar{\cal W}_{-\infty} -  \bar{\cal W}_{\infty} 
\right) \, .
\label{exeqm}
\eeq

Using Eq. (\ref{sptm}) for the superpotential we conclude
that 
\beq
\varepsilon = \frac{4}{\sqrt{3}} (\tilde{m}\tilde\Lambda^5)^{1/2}
\label{vedm}
\eeq
where the scale parameter $\tilde\Lambda$ is restored. The relation 
(\ref{vedm})
holds for any $\tilde{m}$.
It is amusing that, in the limiting case $\tilde{m} \to 0$, half of the 
energy 
density
(\ref{exeqm}) comes from the region $z < z_0$ (a narrow section of 
the wall)
and another half --- from the region $z > z_0$ (a broad section).

When the mass is neither too large, nor too small,  
Eqs. (\ref{eqdw}) do not have analytic solutions. They can be 
solved
 numerically, however. As an illustration the results 
for $\tilde \varphi(z)$, $\tilde 
\chi(z)$
and the parametric plot in the $\varphi,\ \chi$ plane with
$\tilde{m} = 1.0$  are presented on Figs. 2, 3 and 4.
\begin{figure}
\epsfbox{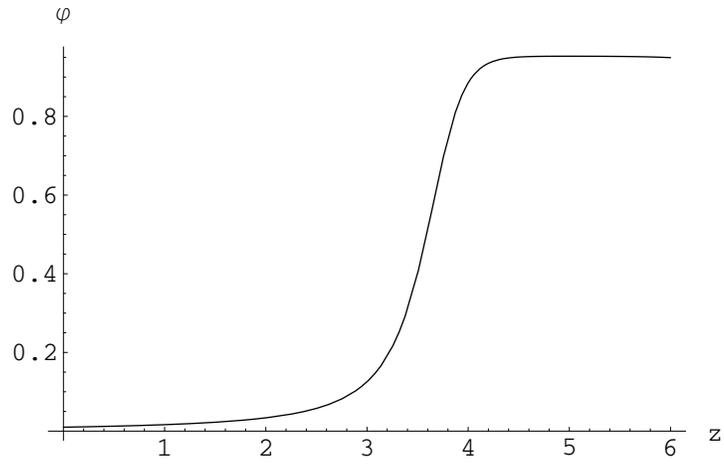}
\caption{ The profile of the field $\varphi$ inside
the domain wall in SUSY QCD for $\tilde{m}=1$.}
\label{f2}
\end{figure}
\begin{figure}
\epsfbox{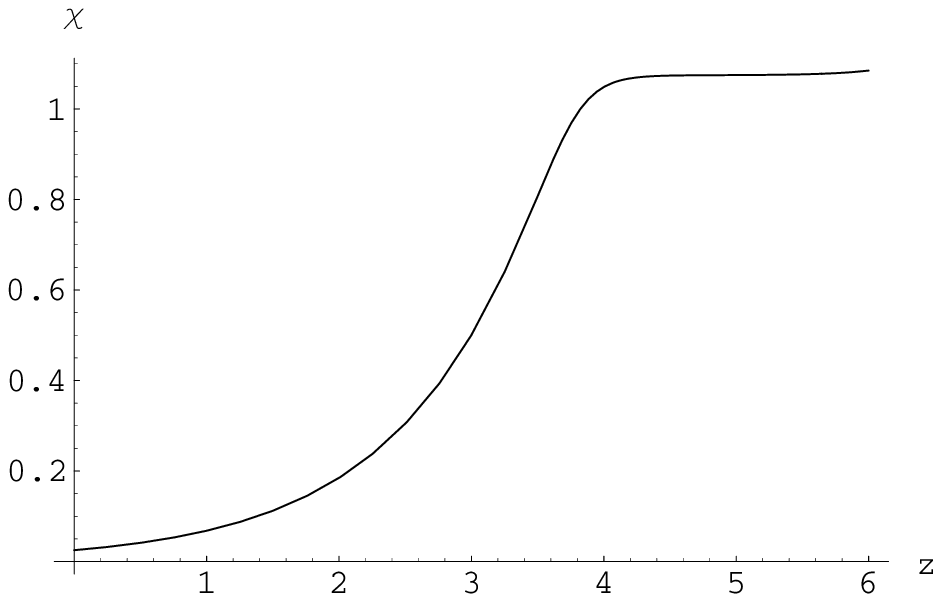}
\caption{ The profile of the field $\chi$ inside
the domain wall in SUSY QCD for $\tilde{m}=1$.}
\label{f3}
\end{figure}

\begin{figure}
\epsfbox{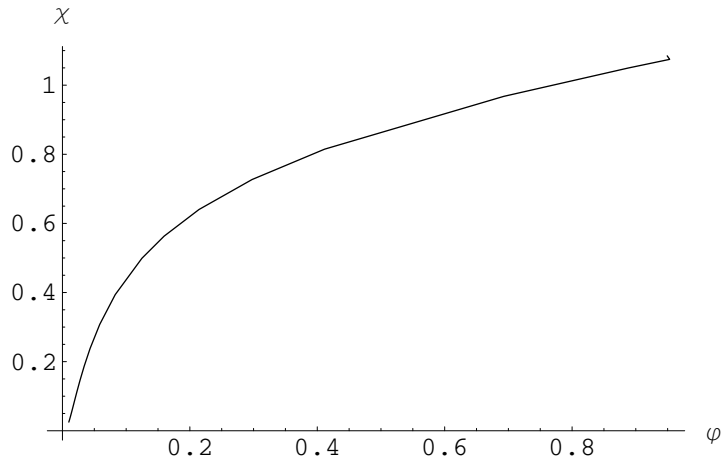}
\caption{ The parametric plot of $\chi$ vs $\varphi$ inside
the domain wall in SUSY QCD for $\tilde {m}=1$.}
\label{f4}
\end{figure}
We solved also the equations for other values of $\tilde{m}$ and 
observed 
how the 
profiles approach the asymptotic curves in the limit
when the mass parameter tends to zero or infinity.

If $\tilde{m}$ is large, the quarks can be integrated out, and SQCD
reduces to supersymmetric gluodynamics. It is not difficult to obtain 
a relation between
the scale parameters
$\tilde\Lambda$ and $\Lambda$. In the fundamental theory this
can be done by exploiting the NSVZ $\beta$ function \cite{NSVZ};
in the effective theories one compares to this end the expressions
for $\langle S\rangle$ in the standard vacuum
in SQCD and supersymmetric gluodynamics, respectively.
In this way we arrive at
\beq
\Lambda^3 = \frac{\sqrt{3}}{2} \left( 
\tilde{m}\tilde\Lambda^5\right)^{1/2}
\, .
\label{LTL}
\eeq
The scale parameters are introduced {\em via} Eqs. (\ref{VYL}) and 
(\ref{VYM}),
respectively. Thanks to holomorphy, the square root
dependence in Eq. (\ref{LTL}) is exact, not approximate.
This relation is exact. Equation (\ref{LTL})  implies, in turn,
that in SQCD $\varepsilon = (4\pi^2)^{-1}$Tr$\lambda^2$ -- this is
exactly the same relation  we observed in supersymmetric 
gluodynamics. The fact that the explicit 
$m$ dependence disappears -- it is completely hidden in the gluino 
condensate -- is no coincidence. The reason is rather transparent:
the central extension of superalgebra (\ref{cext})
in $SU(N)$ SQCD with matter takes the form
\beq
\{ Q^\dagger_{\dot\alpha}Q^\dagger_{\dot\beta}\}
= 4
\left(\vec\sigma\right)_{\dot\alpha\dot\beta}\int
 \, d^3 x \, \vec\nabla  \left\{\sum_{\rm fl}\left[
\frac{m_0}{2} Q^{\alpha f} Q_{\alpha f} + \frac{1}{16\pi^2}
\mbox{Tr}W^2\right]
- \frac{N}{16\pi^2}
\mbox{Tr}W^2 \right\}_{\theta = 0} \, ,
\label{cexm}
\eeq
plus full (super)derivatives. The sum over flavors in the right-hand 
side assumes that
one may have arbitrary number of flavors (the expression above 
refers to  the fundamental representation; we remind that each 
flavor requires two subflavors).
Note that the expression in the square brackets is the Konishi 
anomaly
(\ref{konishi}) itself, and as such, is a full superderivative that gives 
no contribution in the central charge.  Thus, in the theory with any 
number of flavors the central charge is given by the last term in the 
braces, i.e.  
we return to Eq. (\ref{cext}), which holds universally, irrespectively 
of the values of $N_f$ and $m$.

A brief remark is in order here concerning derivation of Eq.
(\ref{cexm}). The commutator consists of two parts.
The first, tree level part is proportional to the matter superpotential
and is entirely due to the matter fields. It 
is trivially obtained from the canonic commutation
relations. The part containing Tr$W^2$ is an anomaly.
In principle, it could be obtained by a direct calculation
of the relevant one-loop graphs, with both the matter and gauge 
fields
in the loop,  provided
these graphs are regularized in the ultraviolet and infrared
in such a way that all symmetries of the model (including 
supersymmetry) are preserved. We did not attempt to carry out this
program in full (although some steps in this direction
are reported in Ref. \cite{CS}).
An indirect way is based on the observation that 
there are no essentially new ``geometric" anomalies other than that
in the divergence of the $R_0$ current and the trace of
the energy-momentum tensor.  A superfield expression for this 
standard ``geometric" anomaly is 
$$
\bar{D}^{\dot{\alpha}}J_{\alpha\dot{\alpha}} = \frac{1}{3} 
D_{\alpha}\left\{
\left[ 3{\cal W } - \sum_i Q_i \frac{\partial{\cal W }}{\partial Q_i}   
\right] - 
\right.
$$
\beq
\left.
\left[ \frac{3N- N_f}{16\pi^2}{\rm Tr}W^2 + 
\frac{1}{8}\sum_i\gamma_i \bar{D}^2
(\bar{Q}_i^+ e^{V} Q_i) \right]
\right\}\, ,
\label{supanom}
\eeq
where $\gamma_i$ are the anomalous dimensions of the quark 
superfields
$Q_i$,
$$
\gamma_i = -\partial\ln Z_i /\partial\ln\mu =
-\frac{N^2-1}{2N}\, \frac{\alpha}{\pi} + ...
$$
(see
 e.g.  Eq. (26) in Ref. \cite{KSV}). 
First, one may discard the last term in the second square brackets,
since this term is a full superderivative.
To make the tree-level part
of this standard anomaly compatible with Eq. (\ref{cexm})
one must subtract from the former full superderivatives
which obviously does not affect the value of the central charge 
anyway. The appropriate superderivative is
 the Konishi relation   (\ref{konishi}) itself. In this way we arrive to 
Eq. (\ref{cexm}), see Ref. \cite{CS}
for further details. An independent check is that at $N_f=0$ Eq. 
(\ref{cexm})
coincides with Eq. (\ref{cext}), which, in turn, can be readily derived,
e.g. from the VY Lagrangian. Equation
(\ref{cexm}) guarantees the smooth transition to the large $m$ limit, 
with no explicit $m$ dependence of the wall energy density.
This feature was anticipated in Ref. \cite{Dvali1}. Another 
independent check is the fact that the anomaly part of the central 
charge (i.e. the coefficient in front of $W^2$) is proportional to
$N_f - N_c$. If we start from $N_f= 0$ and gradually increase this 
parameter we, eventually, come to a point where $N_f =N_c$; here 
the 
coefficient in front of the anomaly part vanishes. The vanishing of 
the anomaly part  at $N_f = N_c$ could have been expected.
Indeed, from Ref. \cite{ADS} we know that at $N_f<N_c$
a non-perturbative superpotential is generated in the massless SQCD, 
while at $N_f = N_c$, although the superpotential could have 
appeared,
it is not generated. In our approach this is due to the vanishing of the 
coefficient in front of $W^2$ in the central charge.

\section{SQCD in the Weak Coupling Regime (Higgs Phase)}

If the mass parameter $m_0$ in Eq. (\ref{matter}) is small
{\it and} the characteristic values of the matter field $M$ are 
assumed
to be large,
the Higgs description of the model is more appropriate \cite{ADS}.
In this case the $S$ superfield can be integrated out,
the supefield $M$ is light (at the classical level
$M$ describes a flat direction if $m_0=0$), and one can obtain
a genuinely Wilsonean  effective Lagrangian for the field $M$ 
(certain restrictions apply in  the domain of small $M$).
We remind that the VY Lagrangians are not Wilsonean
constructions. 

The superpotential of the low-energy theory for the would-be
moduli field $M$ is
\beq
{\cal W} = -\frac{2}{3}\frac{\tilde\Lambda^5}{M} 
-\frac{\tilde{m}}{2}M
\, .
\label{splet}
\eeq
In the fundamental theory it is generated by instantons
\cite{ADS} (for a review see Ref. \cite{VZS}). 
Note, that the superpotential (\ref{splet}) has a discrete
$Z_2$ invariance $M\ra - M$, as it ought to. 
The region
of small $M$ is not legitimate for consideration in this language;
therefore, we will consider the wall solution interpolating between
the vacua with $\langle M \rangle =\pm
(4\tilde\Lambda^5/3\tilde{m})^{1/2} \equiv \pm  \chi_*^2$, 
anticipating 
that
this solution will never go through the region
of small $M$.

The superpotential (\ref{splet}) can be obtained from Eq. 
(\ref{sptm}) by eliminating $S$ through the equation of motion,
$\ln SM = 0$. Note that no trace is left of the two-sector structure
characteristic of the VY Lagrangian in the $SU(2)$ theory. The 
superpotential
for $M$ is perfectly holomorphic. 

Let us again introduce the superfield $X$ such that $ M^2=X^2$ with 
the 
standard kinetic term.
The BPS wall equations corresponding to the superpotential 
(\ref{splet})
for the wall between the vacua $\langle \chi^2\rangle = \pm 
\chi_*^2$ have the 
form
 \beq
\label{eqdwhgs}
\partial_z \chi\ =\ \frac{\partial \bar{\cal W}}{\partial \bar \chi} \ 
=\ 
-\tilde{m}\bar \chi + \frac{4\tilde \Lambda^5}{3\bar \chi^3 }\, ,
  \eeq
and its complex conjugate (there is, of course, also a pair of 
equations
with negative sign with a solution which is the mirror image of the
solution of Eq. (\ref{eqdwhgs})). 
Substituting here $\chi(z) = \rho(z) e^{i\alpha(z)}$,
one observes after some simple transformations that $\rho(z) = 
\chi_*$ is
just constant and the phase $\alpha(z)$ satisfies the equation
 \beq
 \label{alpha}
\partial_z \alpha \ =\ 2\tilde{m}\sin 2\alpha(z)\, .
 \eeq
It is not difficult to solve this equation with the boundary conditions
$\alpha(-\infty) = 0, \alpha(\infty) = \pi/2$. The wall profile thus 
obtained
is 
  \beq
\label{wallhgs}
\chi(z) \ =\ \chi_* \frac{1 + i e^{4\tilde{m}(z-z_0)}}{\sqrt{1 + 
e^{8\tilde{m}(z-z_0)}}}\, .
  \eeq
The energy density of this wall is
\beq
\varepsilon = \frac{8}{\sqrt{3}} (\tilde{m}\tilde\Lambda^5)^{1/2}
\label{vedm1}
\eeq
-- twice the value in (\ref{vedm}). The reason is quite obvious.
The wall (\ref{vedm1}) is not similar to the walls discussed in the 
previous
sections. The latter interpolate between the vacuum
with $\langle\lambda^2\rangle = 0$ and the
standard chirally asymmetric vacuum, 
while the former interpolates between two different 
chirally asymmetric vacua (in the $SU(2)$ theory).

An interesting question is what happens with the wall (\ref{vedm1}) 
in
the case when the ratio $\tilde{m}/\tilde \Lambda$ is not sent to 
zero, but
has a finite value. We do not have an analytic solution in this case. 
But, as was the case for the walls
interpolating between a chirally asymmetric vacuum and
the chirally symmetric one which we discussed in the preceding 
section,
the solution can be found numerically \cite{krug}. Not dwelling on 
details,
 we only mention here that the BPS equations
(\ref{eqdw}) admit 
the wall solutions 
only in some range of masses, $|m| \leq m_* = 4.67059\ldots \tilde 
\Lambda$.
At $\tilde{m} > m_*$, the BPS solution disappear. Moreover, if 
$\tilde{m} > m_{**}
\approx 4.83$, no nontrivial complex wall connecting different 
chirally
asymmetric vacua  exist within the framework of the
VY effective Lagrangian.

\section{Decay of False Vacuum}

Let us now discuss what happens if we
softly break supersymmetry by adding to the Lagrangian a gluino 
mass term
  \beq
 \label{glmass}
\Delta {\cal L}_m \ =\ \frac {m_\lambda} {g_0^2} \left[ {\rm Tr} 
\lambda^2 \ +\ {\rm h.c.}
\right]\, .
  \eeq
Note that in our notation, see Eq. (\ref{SUSYML}), ${\rm 
Tr}\lambda^2$ is a renormalization-group invariant operator, and so 
is the ratio  $m_\lambda/g_0^2$ (to the leading order). We will 
briefly comment on the impact of  subleading terms later. 
It is assumed that $m_\lambda$ is real and positive
(The phase of $m_\lambda$ is equivalent to a $\vartheta$ angle
and is irrelevant).

The degeneracy of the three vacuum states of the
$SU(2)$ supersymmetric gluodynamics  is lifted. We will limit 
ourselves to the effects
linear in the soft supersymmetry breaking. 
In the linear in $m_\lambda$ approximation
the chirally symmetric vacuum stays at zero, the energy density of 
one of the 
chirally asymmetric vacua becomes positive and that of another  
negative,
  \beq
\label{Deps}
{\cal E}_\pm \ =\ \pm \frac {2m_\lambda \Sigma}{g_0^2}
  \eeq
where 
$$
\Sigma  = \langle {\rm Tr} \lambda^2\rangle_+\, ,
$$
and the subscript + in this definition of $\Sigma$
indicates that here  we mean the gluino  condensate in the vacuum 
with the positive value of $\langle {\rm Tr} \lambda^2\rangle$.
In other words, $\Sigma$ is a real positive parameter of dimension
(mass)$^3$. 

Thus, we deal with two false vacuum states that can decay 
into the true vacuum through formation of ``bubbles" \cite{Vol}.

 The decay rate of the false vacuum into the genuine vacuum can be 
 easily evaluated
using the general results of Ref. \cite{Vol}. According to this work, 
the 
decay rate of
the false vacuum is proportional to
  \beq
 \label{rateVol}
 \Gamma \ \propto \ \exp\left\{ - \frac {27}2 \pi^2 \frac 
{\varepsilon^4}{(\Delta {\cal E})^3}
\right\}
  \eeq
where $\Delta {\cal E}$ is the difference of the vacuum energy 
densities in the false and
true vacua, and $\varepsilon$ is the surface energy density of the 
domain 
wall. The estimate 
(\ref{rateVol}) is valid with the exponential accuracy in the ``thin 
wall limit'', i.e.
when the radius of the critical bubble is much larger than the 
characteristic thickness
of the wall or, in other words, when the absolute value of the 
exponent in Eq. (\ref{rateVol})
is large.

To find the decay rate of the chirally symmetric vacuum in the true 
vacuum with the negative
energy density (the one with positive energy density would decay in 
two 
stages), we have to substitute
in Eq. (\ref{rateVol}) the expression (\ref{Deps}) for $\Delta {\cal E}$ 
and the expression
(\ref{vacen}) for $\varepsilon$ (assuming $N=2$, $\langle 
\mbox{Tr}\, 
\lambda^2\rangle_\infty
= \Sigma$ and $\langle \mbox{Tr}\, \lambda^2\rangle_{-\infty} = 
0$). We then obtain
 \beq
\label{rateSYM}
\Gamma \ \propto \ \exp\left\{- \frac{27}{4096\pi^6} \frac 
{\Sigma}{(m_\lambda/g_0^2)^3}\, .
 \right\}
\label{fvd}
 \eeq
Note a rather small numerical factor in the exponent. The 
quasiclassical formula 
(\ref{rateSYM})
 is valid when $m_\lambda/g_0^2 \ll 0.02 \, \Sigma^{1/3}$. It 
applies  both to supersymmetric
 gluodynamics and to SQCD.

If the gluino mass term has a phase (or if $\vartheta\neq 0$, which 
is 
the same),
$$
m_\lambda  \ra |m_\lambda  |e^{i\alpha}\, ,
$$
the parameter $m_\lambda $ in the exponent of Eq. (\ref{fvd})
is substituted by
$$
|m_\lambda  \cos\alpha |\, .
$$
At $\alpha =\pi/2$ the exponent  becomes infinite.
The reason is obvious: for purely imaginary gluino mass term the
vacuum degeneracy is not lifted, and there is no false vacuum decay. 

The false vacuum decay rate is a physical quantity, and as such
it must be independent of the normalization point $\mu$.
It was already mentioned that $\Sigma$ is renormalization-group 
(RG) invariant. The ratio $m_\lambda/g_0^2$ is RG invariant only in 
the leading logarithmic approximation. Beyond the leading 
approximation the exact RG invariant combination is \cite{HS}
\beq
m_\lambda \left(\frac{1}{g_0^2} - \frac{1}{4\pi^2}\right)\, .
\label{RGI}
\eeq
Therefore, $m_\lambda /g_0^2$ in the exponent in Eq. (\ref{fvd})
is actually substituted by the combination (\ref{RGI}), see \cite{HS}
for further details.

\section{Walls vs. Torons.}

All the previous discussion was based on the assumption that the
topological charge can only be an integer.
There is a lasting controversy in the
literature as to the question of existence of configurations
with fractional topological charge in pure glue $SU(N)$ gauge
theories.

In supersymmetric $SU(N)$ Yang-Mills theories 
the question of existence (or non-existence) of the domain walls 
interpolating between different chirally asymmetric vacua 
is in one-to-one correspondence with the question of the proper 
quantization of
the topological charge (\ref{topch}). If the minimal non-trivial 
topological charge is unity, the presence of $N$ different vacuum 
states
implies the spontaneous breaking of the physical discrete symmetry
$Z_{2N} \to Z_2$ and, correspondingly, the appearance of domain 
walls.
An alternative, with a new superselection rule replacing
the physical $Z_{2N}$ invariance, arises if fractional values of the 
topological charge are possible. Let us elucidate this assertion in 
more 
detail. 

If the $SU(N)$ Yang-Mills theory is compactified on 
 four-dimensional 
torus, with a finite size $L$, 
the topological charge is quantized  fractionally. The so 
called
toron field configurations, with $\nu$ being multiple integer of 
$1/N_c$,
do exist \cite{Hooft}. An {\it assumption} that such configurations 
survive and contribute
in the path integral  in the large-volume limit $L\ra\infty$ leads to 
the 
conclusion
that the $Z_N$ transformation connects vacua with diffferent
$\vartheta$ rather than physically distinct  degenerate vacua. If 
large
gauge transformations can change the winding number of the 
gauge field
configuration by $1/N_c$, while the vacuum angle $\vartheta$ is 
defined in the ``old" way, see Eq. (\ref{SUSYML}), then the vacuum 
angle $\vartheta$ varies within the 
range
$0 \leq \vartheta \leq 2\pi N_c$, and the sectors with different 
$\vartheta$ 
do not
communicate with each other. A new  superselection 
rule
should be imposed. Alternatively one may say that a new
vacuum angle should be defined, $\tilde\vartheta = N_c^{-1} 
\vartheta$. Then $\tilde\vartheta$ would vary between zero and 
$2\pi$, as is appropriate for the vacuum angle. $N_c$ chirally 
asymmetric vacua of supersymmetric
gluodynamics with $\langle\mbox{Tr}\lambda^2\rangle
\propto\exp (2\pi i k/N_c)$ would correspond to different values of 
$\tilde\vartheta =0, 2\pi/N_c, 4\pi/N_c, ...$ and could not coexist in 
one 
and the same Universe,
 (see \cite{Leut,SMtor} for a detailed 
discussion).
If it were true, one could not speak of domain walls between the 
different
supersymmetric vacua. In particular, the 
expression
(\ref{fvd}) for the decay rate of the metastable vacuum in a theory 
where 
supersymmetry
is slightly broken would not make sense.

The most serious argument in favor of this viewpoint 
comes from 
the calculation of the gluino condensate on the small torus 
\cite{Gomez}. In the $SU(N)$ supersymmetric gluodynamics 
the gluino condensate turns out to be saturated by the toron field 
configurations. The 
expression
has the form
   \beq
\label{condtor}
   \langle {\rm Tr} \lambda^2\rangle  \sim  \frac{1}{L^3g^2(L)} \exp 
\left\{ - \frac{8\pi^2}
{N g^2(L)} \right\}
  \eeq
where $L \ll \Lambda^{-1}$. Using the exact NSVZ $\beta$ function
\cite{NSVZ}
we find that the condensate actually does not  depend 
on $L$; the toron result is equal to a numerical constant times
 $\Lambda^3$. From this one could tentatively 
conclude that
the toron contribution to the condensate survives for large $L$ 
\cite{Zhit,Leut}. 

Clearly, if we allow the toron configurations, the $Z_N$ 
as a physical symmetry disappears
from the VY Lagrangian. Say, in the $SU(2)$ theory, Eq. (\ref{VYL}) 
must be modified: $2\pi i n$ on the right-hand side must be replaced 
by $4\pi i n$. Correspondingly, instead of two sectors of the scalar 
potential (see Eq. (\ref{scalpot}) and the following discussion)
we will have just one sector extending in the interval
arg$\phi \in (-\pi, \pi)$. This scalar potential would have only two 
minima: one chirally asymmetric at $\phi =1$, and the second 
chirally symmetric
minimum at $\phi =0$.

Now, the argument in favor of torons is not free of
 inconsistencies (see \cite{QCD2} for a 
recent detailed discussion).
 First, the existence of the field configurations with 
fractional 
topological charge relies on the fact that the theory is compactified 
on the 
torus. If the theory is compactified on $S^4$, only the field 
configurations
with integer topological charge are admissible and there are no 
of torons. 
But the physics should not depend on whether the theory is 
compactified on a sphere,
on a torus or on some other manifold, if the size of the manifold is 
much 
larger than
the characteristic scale $\Lambda^{-1}$.

Second, in the theories with higher orthogonal and exceptional 
groups, configurations
with fractional topological charge do not exist even on  torus, and it 
is 
{\it not} possible to interpret the vacuum degeneracy in these 
theories in the
language of new superselection rules.

Third, if the matter fields in the fundamental representation are
introduced, torons disappear since the twisted boundary conditions 
on 
a torus necessary for their existence cannot be imposed. When the 
matter
mass term tends to infinity we return back to supersymmetric 
gluodynamics.
If the transition is smooth, and we have seen that, for the walls 
connecting 
a chirally asymmetric vacuum and the chirally symmetric one. it {\it 
is}
smooth, the energy
density  of the domain wall remains finite. Then the torons must
be irrelevant in supersymmetric gluodynamics too.

Finally, the exact solution of $N=2$ supersymmetric gluodynamics 
found recently \cite{SieWit},  shows no traces of the presence of 
torons and fractional topological charges. Although this solution
is not rigorously proven, it is perfectly self-consistent
and goes through numerous indirect checks. 

Our present viewpoint is that the domain walls are real -- they {\it 
do}  exist
in supersymmetric gluodynamics and SQCD. The cleanest
 argument comes from the consideration of the theory with 
matter in the weak
coupling regime $\tilde{m} \ll \tilde\Lambda$. In this case the  
vacua 
$\chi^2 = 
2/\sqrt{3\tilde{m}}$
and $\chi^2 = - 2/\sqrt{3\tilde{m}}$ acquire large classical Higgs 
vacuum 
average, and 
the walls
separating them exist by the same token as in the most trivial model 
of 
one  real
scalar field with the double well potential 
$V(\phi) \ =\ \lambda(\phi^2 - v^2)^2$. 

In the theory with fundamental matter there is no place for 
an argument:  field configurations with fractional topological 
charges do not
exist no matter how the theory is compactified
since the twisted boundary
conditions on a torus necessary for their existence cannot be 
consistently
imposed. A remarkable corollary of supersymmetry is the
holomorphic dependence (\ref{vedm1}) of the wall energy density 
on the
mass. As was explained before, the energy density 
(expressed via the physical scale (\ref{LTL}) ) remains
finite in the limit
$m \to \infty$ when the matter fields decouple and we are left with 
the
pure supersymmetric gluodynamics.   This means that in this case, 
the proper topological classification should involve only integer
topological charges.

It is very instructive to confront this situation with
a simple 
(non-supersymmetric) two-dimensional model where a similar 
question can be posed
and exactly answered, but the answer is precisely opposite. 
Consider two-dimensional QED (the Schwinger 
model) with 
two fermion flavors,
a massless fermion $\psi$ with the charge $e$ and a massive fermion 
$\Psi$
 with the charge $e/2$:
  \beq
 \label{LSMferm}
{\cal L} \ =\ -\frac 14 F_{\mu\nu} F_{\mu\nu} \ +\ \bar \psi 
(i\gamma_\mu \partial_\mu
- e \gamma_\mu A_\mu) \psi \ + \ \bar \Psi \left(i\gamma_\mu 
\partial_\mu
- \frac e2  \gamma_\mu A_\mu - M\right) \Psi 
 \eeq
where 
$$
F_{\mu\nu} = \ \partial_\mu A_\nu - \partial_\nu A_\mu \ 
\equiv \epsilon_{\mu\nu} F\, .
$$

In the Euclidean version of the theory, the topological charge
  \beq
  \label{nu2}
\nu_2 \ =\ \frac {e/2}{2\pi} \int F(x) d^2x \ =\ Z
 \eeq
is quantized. The minimal flux is determined by the boundary 
condition on the field $\Psi$, i.e. by the charge $e/2$ of 
the heavy fermion
field. The minimal topological charge is unity.

In the limit $M \to \infty$, heavy fermions decouple. We must 
consider now only the boundary conditions on the field $\psi$.  
The gauge field 
configurations with half-integer flux $\nu_2$ become allowed. This 
simulates the
situation in four-dimensional  supersymmetric gauge theories with 
and without fundamental matter. When SQCD 
involves dynamical
fields in the fundamental representation, only integer topological 
charges 
(\ref{topch}) are admissible. In the limit $m \to \infty$, we are left 
 with
the adjoint gauge fields and adjoint fermions (gluinos), and, on a 
torus, 
gauge field configurations with 
fractional
topological charge show up.

Let us now address the question of ``domain walls'' in our 
two-dimensional model.
Quotation marks are used above to remind the reader that, due to 
the lack of extra two dimensions 
these
objects are not really ``walls'', but rather localized soliton 
configurations. 
The existence
of such solitons was recently discovered in \cite{Kleb}. It is best seen 
by bosonizing
the theory according to the rules \cite{Luther,boson}
  \beq
  \label{boson}
\bar \psi \gamma_\mu \psi \ \to \  \frac 1{\sqrt{\pi}} 
\epsilon_{\mu\nu} \partial_\nu 
\phi,\ \ \ \ \ 
\bar \Psi \gamma_\mu \Psi \ \to \  \frac 1{\sqrt{\pi}} 
\epsilon_{\mu\nu} \partial_\nu 
\chi \, ,\nonumber \\
\bar \psi \psi \ \to \ -\mu_\psi \cos (2\sqrt{\pi} \phi), \ \ \ \ \ \ 
\bar \Psi \Psi \ \to \ -\mu_\Psi \cos (2\sqrt{\pi} \chi)\, ,
  \eeq
where the constants $\mu$ carrying the dimension of mass depend 
on a particular
normalization procedure for the scalar fermion bilinears. Assuming 
$M \gg e$ and
integrating out the gauge fields, one arrives at the bosonized 
Lagrangian involving 
only the 
physical degrees of freedom  
  \beq
  \label{LSMbos}
{\cal L}_{\rm bos} \ =\ \frac 12 (\partial_\mu \phi)^2 + 
\frac 12 (\partial_\mu \chi)^2\ -\  \frac{e^2}{2\pi} \left( \phi + \frac 
\chi 2 \right)^2
\ +\ cM^2 [\cos (2\sqrt{\pi} \chi) - 1]\, .
 \eeq
In contrast to the four-dimensional case where the effective boson 
Lagrangians are
approximate, here the Lagrangian (\ref{LSMbos}) is {\it exactly} 
equivalent to
(\ref{LSMferm}) in the sense that the spectrum and all other 
physical 
properties of
the theories (\ref{LSMferm}) and  (\ref{LSMbos})  
coincide\footnote{Generally speaking, the Lagrangian (\ref{LSMbos}) 
is not 
quite correct.
As in the case of the standard Schwinger model \cite{QCD2}
we have to separate the zero spatial Fourier harmonics of the 
``massive photon''
field  $\phi + \chi /2$
 and write down a ``glued" 
Lagrangian similar to that in SUSY gluodynamics, 
 invariant under the transformation $\phi + \chi/2 \ \to\ 
\phi + \chi/2 \ +\ \sqrt{\pi}/2$. This is irrelevant
for the discussion which follows:  the 
combination $\phi + \chi /2$ inside the wall does not cross the 
boundary,
$|\phi + \chi /2|$ is always less than $\sqrt{\pi}/2$.}.

We see that the
potential in  Eq. (\ref{LSMbos}) involves an infinite set of minima at
$\chi \ =\ -2\phi \ = \ n\sqrt{\pi}$ with integer $n$. The theory 
admits finite energy 
static solutions which interpolate between $\phi = \chi = 0$ at $x = 
-\infty$
and, say, $\chi = -2\phi = \sqrt{\pi}$ at $x = \infty$. The physical 
meaning
of such a soliton is clear. It is a kind of a ``heavy meson''
(cf. Ref. \cite{Karl} where such  meson solutions were
obtained in a theory where  the charges of the light and heavy fields 
were equal) 
composed of the 
original heavy quark $\Psi$ and a cloud of massless fermion fields 
$\psi$ which
neutralize its charge (a ``constituent quark'' in  terminology of Ref.
\cite{Karl}).
The fact that integer-charged light fermions manage to screen a 
heavy fractional
charge \cite{Jackiw,Kleb} crucially depends on  
the masslessness of 
the light fermion. If $m_\psi \neq 0$, the energy of such a 
``meson'' becomes infinite and heavy fractional charges are 
confined.

In the context of our discussion here, two facts are important.
 The  light quark condensate $\langle \bar \psi \psi\rangle \ 
\propto \
\langle \cos(2\sqrt{\pi} \phi)\rangle$ changes sign in passing from 
$x = -\infty$ 
to $x = \infty$
along the soliton solution. Thereby, these solitons are very much 
analogous
to the four-dimensional supersymmetric domain walls separating 
different 
vacua. Indeed, the 
Lagrangian (\ref{LSMbos})
involves a discrete $Z_2$ symmetry corresponding to the positive or 
negative
sign of  the light quark condensate. This 
symmetry is broken
spontaneously\footnote{For those interested in the in-depth 
coverage we note that the spontaneous breaking takes place only at  
zero temperature. At any nonzero temperature, the ``domain
walls'' which mix the distinct vacua  appear in the heat bath and the 
symmetry is
restored. The density of the soliton states is $\propto \ \exp\{-
E_{\rm sol}/T\}$.
 The ``domain walls'' have
finite energy here due to the absence of extra transverse 
dimensions. The situation is 
exactly the same as in the one-dimensional Ising
model (see e.g. \cite{Baxter}) and in $QCD_2$ with adjoint fermions 
for higher
unitary groups \cite{QCD2}: in all these cases we have a first order 
phase transition at $T=0$.}. 

 Now, we come to a crucial distinction of this two--dimensional model
 as compared to the four--dimensional supersymmetric theory. 
In the two-dimensional model at hand, 
 when $M$ is large, the energy of the soliton is of  order of the 
mass of the heavy
fermion $M$ (the ``constituent quark'' has an energy of order of  $e 
\ll M$). This means 
that, in the limit $M \to \infty$, the mass of solitons
becomes infinite. Correspondingly, the sectors with different signs of  
$\langle \bar \psi 
\psi\rangle $ cease to
talk to each other, which nicely conforms with the standard 
topological classification
of the  Schwinger model with  one dynamical massless fermion 
of charge $e$, 
 where the flux (\ref{nu2}) (a half of
the standard flux) is quantized to half-integer values.

As was repeatedly mentioned, in the four-dimensional $SU(2)$
supersymmetric Yang--Mills theory, 
the walls interpolating between
chirally symmetric and asymmetric vacua  are BPS-saturated.
 By virtue of the  exact
theorem (\ref{cext}) and its corollary (\ref{vacen}) that guarantees 
that
their energy density remains finite also in the limit 
$m\to \infty$.
By combining two walls of the type we found 
-- one interpolating between 1 and 0, and another between
0 and $- 1$  -- we may build a wall interpolating
between two chirally asymmetric vacua in the
$SU(2)$ supersymmetric gluodynamics. This wall just consists
of two independent components. This ``superposition" of two 
BPS-saturated
solutions in the
$SU(2)$ supersymmetric gluodynamics, going through $\phi = 0$
along the real axis is almost a  BPS-saturated solution
in itself. That is the energy of this configuration approaches the BPS 
bound
when the two components of the wall are far apart.
\footnote{For small masses, there is also another kind of domain 
walls
 (\ref{wallhgs}) connecting 
different chirally asymmetric vacua directly, without passing 
through zero.
We know now that {\it such} walls disappear at large enough masses 
\cite{krug},
but this fact seems to be irrelevant for the present discussion.}
For $SU(3)$ and higher groups
superimposing two solutions (one goes from 1 to 0, 
and another goes from 0 to ${\rm e}^{2\pi i/ N}$ 
along the straight lines in the complex $\phi$ plane) also
gives a 
two-component domain wall with the interacting components.
Stricktly speaking,
this configuration is not a solution of equations of motion, 
although it approaches
a solution in the limit of infinitely large separations
between the components of such a wall.
The truely BPS-saturated
walls, if they exist in this case, should go
through the complex $\phi$ plane in a non-trivial manner.

Returning to the torons, our
consideration implies that, for large 
volumes,
the relevant topological classification in SUSY gluodynamics 
is exactly the same as in $SQCD$ 
-- {\it the topological charge 
(\ref{topch})
is strictly integer and there are no torons}.

The arguments presented here, although 
rather
convincing to our mind, are unfortunately indirect.
To resolve the paradox completely one should explain why 
the toron 
configurations  which are essential for small tori, see Eq. 
(\ref{condtor}),  disappear in the 
limit
$L \to \infty$. We think that that's very much probable, but 
at the moment do not 
 see a
technical reason for it.

\section{Conclusions}
In this work we have studied domain walls in supersymmetric 
gluodynamics
and in SQCD. There are two basic types of walls in these theories: a 
wall
that interpolates between two chirally asymmetric vacua and a wall
that interpolates between a chirally asymmetric vacuum and a 
symmetric vacuum
at the origin of the field space.
We have shown explicitly that both those types of walls are BPS 
saturated.
The energy density of BPS saturated walls in supersymmetric 
theories 
is unambiguosly determined by the difference between the
asymptotic values of the chiral condensate
on the two sides of the wall.
We have calculated this energy density explicitly from 
the (corrected) Veneziano-Yankielowicz effective Lagrangean and 
found that it indeed
conforms with the exact relation (\ref{vacen}).

When a small supersymmetry breaking mass is added to the SUSY 
gluodynamics
Lagrangean, the degeneracy between the vacua is lifted. All vacuum 
states
except for one become metastable. We have calculated the decay rate 
of these
false vacua to leading order in the SUSY breaking mass.
It is an interesting question wether some of these metastable states 
survive
at large mass and therefore exist also in pure nonsupersymmetric 
Yang-Mills
theory.

Although the explicit calculations in this paper where performed for 
the
gauge group $SU(2)$, the exact relation (\ref{vacen}) is valid for any
$SU(N)$ group and we do not expect any qualitative changes in the
character of the wall solutions for higher $N$.

\vspace{0.5cm}

{\bf Acknowledgments}: \hspace{0.2cm} 
M.S. is grateful to M. Green and P. Van Baal
for useful discussions. A.S. acknowledges warm hospitality extended
to him at 
Theoretical Physics Institute, University of Minnesota.

This work was supported in part by DOE under the grant number
DE-FG02-94ER40823, by the INTAS grants 93--0283 and 94--2851,
by the U.S. Civilian Research and Development Foundation under 
award 
\# RP2--132, by the Schweizerishcher National 
Fonds grant \# 7SUPJ048716 and by the RFFI grant 97-02-16131.

\end{document}